\documentclass{article}
\usepackage{amsfonts,amssymb, amsmath,mathrsfs}

\usepackage[english]{babel}

\def\nn{\nonumber }
\def\bq{ \begin{equation} }
\def\eq{ \end{equation} }
\def\ben{ \begin{eqnarray} }
\def\en{ \end{eqnarray} }

\def\g{{\gamma}}

\newtheorem{prop}{Proposition}

\newtheorem{re}{Remark}
\newenvironment{rem}{\begin{re} \rm }{\end{re}}
\begin{document}


\title{On the Lie integrability theorem for the Chaplygin ball}
\author{ A V Tsiganov \\
\it\small
St.Petersburg State University, St.Petersburg, Russia\\
\it\small e--mail: andrey.tsiganov@gmail.com}

\date{}
\maketitle

\begin{abstract}
The necessary number of  commuting vector fields for the Chaplygin ball in the absolute space is constructed. We propose to get these vector fields in framework of the Poisson geometry similar to  the Hamiltonian mechanics.
\end{abstract}

\section{Introduction}
\setcounter{equation}{0}

As in \cite{ch03} we consider the rolling of a dynamically balanced ball on  horizontal absolutely rough table without slipping or sliding. `Dynamically balanced' means that the geometric center coincides with the center of mass, but the mass distribution is not assumed to be homogeneous. Because of the roughness of the table this ball cannot slip, but it can turn about the vertical axis without violating the constraints.

Let $M=(M_1,M_2,M_3)$ be the angular momentum of the ball with respect to the contact point and $\omega=(\omega_1,\omega_2,\omega_3)$ be the angular velocity vector of the rolling ball. Its mass, inertia tensor and radius will be denoted by $m$, $\mathbf I = \mathrm{diag}(I_1, I_2, I_3 )$ and $b$ respectively. Here and below all the vectors are expressed in the so-called body frame, which is firmly attached to the ball, and its axes coincide with the principal inertia axes of the ball.

Equations of motion
\bq\label{m-eq}
 \dot \alpha=\alpha\times \omega\,,\qquad \dot \beta=\beta\times \omega\,,\qquad \dot \g=\g\times \omega\,, \qquad \dot M=M\times \omega\,,
\eq
have the same form as the Euler-Poisson equations in rigid body dynamics \cite{ch03,bm01,kil02}.
In fact the principal difference between holonomic and nonholonomic systems is chidden within the relation of the angular velocity with the angular momentum
\bq\label{omega}
 \omega=\mathbf A_\g M\,,
\eq
where in the Chaplygin case
\bq\label{ch-sol}
 \mathbf A_\g=\mathbf A +\dfrac{d\,\mathbf A\,\g \otimes\g\,\mathbf A}{g^{2}(\g)}\,,\qquad
 \mathbf A=\left(
 \begin{array}{ccc}
 a_1 & 0 & 0 \\
 0 & a_2 & 0 \\
 0 & 0 & a_3
 \end{array}
 \right),\qquad a_k=(I_k+d)^{-1}\,,
\eq
and
\bq\label{g-fun}
g(\g)=\sqrt{1-d(\g,\mathbf A\g)}\,,\qquad\qquad d=mb^2\,.
\eq
Here and below we always assume that unit vector $\g$ is perpendicular to the plane.
 \begin{rem}
 At $d=0$ one gets standard Euler-Poisson equations for the Euler-Poinsot top, which is a well-studied  holonomic Hamiltonian dynamical system.
\end{rem}
According to the kinematic interpretation, the first three equations in (\ref{m-eq}) describe the evolution of a fixed in the space ortonormal inertial frame rotating with the angular velocity $\omega$. Basis unit vectors
\[\alpha=(\alpha_1,\alpha_2,\alpha_3), \qquad \beta=(\beta_1,\beta_2,\beta_3),\qquad \g=(\g_1,\g_2,\g_3)\]
of the inertial frame expressed in the rotating reference frame are columns of the real orthogonal rotation matrix
\begin{equation}\label{10}
\mathbf R=
\begin{pmatrix}
\alpha _1&\beta_1&\gamma _1\\
\alpha _2&\beta_2&\gamma _2\\
\alpha _3&\beta_3&\gamma _3\\
\end{pmatrix}
\in SO(3),
\end{equation}
Consequently vector field $X$ defined by (\ref{m-eq}) has six geometric or kinematic integrals of motion
\bq\label{c-int}
\begin{array}{lll}
C_1=(\alpha,\alpha)\,,\qquad &C_2=(\beta,\beta)\,,\qquad &C_3=(\g,\g)\,,\\
C_4=(\alpha,\beta)\,,\qquad &C_5=(\alpha,\g)\,,\qquad &C_6=(\beta,\g)\,,
\end{array}
\eq
which are equal to one ($C_1,\ldots,C_3$) or to zero ($C_4,\ldots,C_6$).

Because  ball is a "free" rigid body not subject to any external force, spatial components of the angular momentum
\bq\label{j-int}
J_1=(\alpha,M)\,,\qquad J_2=(\beta,M)\,,\qquad J_3=(\g,M)
\eq
are three integrals of motion and replacing one of them by the sum of their squares yields a second quadratic in momenta integral of motion
\[
M^2=M_1^2+M_2^2+M_3^2=J_1^2+J_2^2+J_3^2
\]
as well as  the kinetic energy
\bq\label{ham}
 H=\dfrac12\,(M,\omega)\,.
\eq
According to \cite{ch03} vector field $X$ has an invariant volume form
\bq\label{volume-form}
\Omega=g^{-1}(\g)\,\mathrm d \alpha\wedge\mathrm d\beta\wedge\mathrm d\gamma\wedge\mathrm d M\,.
\eq
Thus, according to the Euler-Jacobi theorem it is possible to integrate the system (\ref{m-eq}) by quad\-ra\-tu\-res.

The main aim of this paper is to prove the integrability of the vector field $ X $ (\ref{m-eq}) by means of  the Lie integrability theorem.

\subsection{The Euler-Jacobi and Lie integrability theorems}
Let us remind of some facts about various  integrability theorems following  the recent work \cite{koz13}.

According to Euler and Jacobi, vector field $X$ defined by
\bq\label{m-eq2}\dot{x}_i = X_i(x_1,\dots, x_n)\eq
is integrable by quadratures if there are $n-2$ functionally independent first integrals $f_1,\dots , f_{n-2}$ and an invariant volume form
\bq\label{mu-gen}
\Omega = \mu\, \mathrm dx_1\wedge \mathrm dx_2\wedge \dots \wedge \mathrm dx_n\,.
\eq
In the classical mechanics this assertion is often referred to as the Euler-Jacobi theorem of the last multiplier.
In this case the compact connected component of the level set of integrals of motion without singular points is a smooth manifold diffeomorphic to the two-dimensional torus, see \cite{arn}.

A classical Lie theorem \cite{lie} says that the vector field $X$ (\ref{m-eq2}) is integrable by quadratures if there are $n$ independent
vector fields
\bq\label{u-lie}
u_1=X,\, u_2,\ldots, u_{n}
\eq
which generate a solvable Lie algebra with respect to the commutation operation $[. ,. ] $, modern discussion may be found in \cite{koz13,olver00}.

The Jacobi last multiplier $\mu$ (\ref{mu-gen}) provides an incestuous interrelation between symmetries,
first integrals and integrating factors for well-endowed systems, see Lie \cite{lie}, page 343-347 and Bianchi \cite{bian} pages 456-464.

For instance, if we know all $n$ generators (\ref{u-lie}) of the symmetry algebra
\[
u_i=\sum_{j=1}^n a_{ij}\partial_{x_j}\,,\qquad i=1,\ldots,n\,,
\]
the Jacobi last multiplier (\ref{mu-gen})  is also given by
\bq\label{mu-det}
\mu=g^{-1}\,,\qquad g=\det\left[
 \begin{array}{cccc}
 a_{11} & a_{12} & \cdots & a_{1n} \\
 \vdots & & & \vdots \\
 a _{n1} & a_{n2} &\cdots & a_{nn} \\
 \end{array}
 \right]\,,
\eq
in  case of $g\neq 0$. So, in the classical Lie integrability theorem we usually have $n+1$ tensor invariants.

In generic cases the vector field $X$ (\ref{m-eq2}) is integrable by quadratures if there are $k$ functional independent integrals of motion
\bq\label{f-gen} f_1\,,\ldots, f_k
\eq
and $n-k$ linearly independent vector fields
\bq\label{u-gen}
u_1=X,\, u_2,\ldots, u_{n-k}
\eq
which generate a solvable Lie algebra, and the Lie derivatives along any of the fields
 $ u_i $ from any of the integrals $ f_j $ are equal to zero
\bq\label{l-gen}
\mathcal L_{u_i}\,f_j=0\,.
\eq
The above statement follows immediately from the Lie theorem applied to the
restriction of the dynamical system (\ref{m-eq2}) on the integral manifold
\[
\mathcal N_c=\{x\in \mathcal M\,,\qquad f_1=c_1\,,\ldots,f_k=c_k\}\,.
\]

In fact, the Lie integrability theorem obviously implies the well-known Liouville theorem
on the integrability of the Hamiltonian vector field with
 a complete set of independent integrals of motion in involution. In this simplest case
 $n = 2k$, vector field $X$ (\ref{m-eq2}) is the Hamiltonian vector field
 \[
 u_1=X=P\mathrm dH\,,\qquad H=f_1\,,
 \]
 and $k$ functional independent integrals of motion $f_j$ (\ref{f-gen}) are in  involution
 \bq\label{f-ham}
 \{f_i,f_j\}=\sum_{lm=1}^n P_{lm}\dfrac{\partial f_i}{\partial x_l} \dfrac{\partial f_j}{\partial x_m}=0\,,\qquad i,j=1,\ldots,k,
 \eq
 with respect to the Poisson brackets defined by the Poisson bivector $P$.
 So $k$ vector fields
 \bq\label{u-ham}
 u_i=P\mathrm df_i
 \eq
 are linearly independent at each point, they commute pairwise and relations (\ref{l-gen}) hold true. See \cite{arn,van11,neh05} for a more comprehensive discussion on integrability theorems and related issues.

Two important consequences of the Jacobi identity for $P$ are that the generalized dis\-tri\-bu\-tion on $\mathcal M$, defined by the Hamiltonian vector $X=PdH$, is integrable, and that the Hamiltonian vector fields (\ref{u-ham}) which are associated with Poisson commuting functions (\ref{f-ham}) are commuting vector fields.

\section{Linear Poisson bivectors for the Chaplygin ball}
\setcounter{equation}{0}

In the holonomic case at $d=0$ the vector field $X$ (\ref{m-eq}) is a Hamiltonian vector field
\[X=P\mathrm d H\,,\]
with respect to the Hamiltonian (\ref{ham}) and to the canonical Poisson bivector
\bq\label{can-poi}
P=\left(
 \begin{array}{cccc}
 0& 0 & 0& \boldsymbol{A} \\
 0 & 0 &0 & \boldsymbol{B} \\
 0 & 0 & 0 & \boldsymbol{\Gamma} \\
 \boldsymbol{A} & \boldsymbol{B} & \boldsymbol{\Gamma} & \boldsymbol{M} \\
 \end{array}
 \right)\,.
\eq
Here we use a convention that any vector $z$ is identified with the $3\times3$ skew-symmetric matrix $\boldsymbol{Z}$
\[
z=\left(
 \begin{array}{c}
 z_1 \\
 z_2 \\
 z_3 \\
 \end{array}
\right)\to \boldsymbol Z=\left(
 \begin{array}{ccc}
 0 & z_3 & -z_2\\
 -z_3 & 0 & z_1 \\
 z_2 & -z_1 & 0 \\
 \end{array}
 \right)\,.
\]
\begin{rem}
At $d=0$ it is easy to see that the following change of variables
\[
\g\to\g=\dfrac{\g\times M}{|\g\times M|}
\]
preserves the equations of motion (\ref{m-eq}) and reduces the linear in momenta Poisson bivector $P$ (\ref{can-poi}) to a rational Poisson bivector. In order to avoid such freedom below we will consider only linear in momenta Poisson bivectors.
\end{rem}

Integrals of motion (\ref{c-int}) are the Casimir functions of $P$ (\ref{can-poi})
\[P\mathrm d C_k=0\,,\]
which give rise to the trivial vector fields. Spatial components of the angular momentum (\ref{j-int})
generate $so(3)$ algebra
\bq\label{j-br}
\{J_i,J_j\}=\varepsilon_{ijk} J_k\,,
\eq
where $\varepsilon_{ijk}$ is a totally skew-symmetric tensor. The two remaining integrals of motion $H$ and $M^2=(M,M)$ are in involution with all the other integrals .

Thus, in the holonomic Euler case at $d=0$ we have nine functionally independent integrals of motion
\[C_1,\dots,C_6\,,\qquad f_1=H\,,\qquad f_2= M^2\,,\qquad f_3= (\g,M)\,,
\]
and three commuting to each other vector fields
\[
u_1=P\mathrm df_1\,,\qquad u_2=P\mathrm d f_2\,,\qquad u_3=P\mathrm d f_3\,,
\]
so that
\[\mathcal L_{u_i} f_j = 0 \,,\qquad \forall i,j\,.\]
It allows us to say that vector field $X$ can be integrable by quadratures according to the Lie integrability theorem.

\subsection{Nonholonomic case}
As in \cite{ch03} let us start with the projection of the initial vector field $X$ (\ref{m-eq})
 \[
 \hat{X}:\qquad \dot \g=\g\times \omega\,, \qquad \dot M=M\times \omega
 \]
 on the six-dimensional phase space $\hat{\mathcal M}$ with coordinates $\g$ and $M$.
According to \cite{bm01} the vector field $\hat{X}$ is a conformally Hamiltonian vector field
\[
\hat{X}={g}^{-1}\, \hat{P}_g\, \mathrm dH
\]
with respect to the Poisson bivector
\bq\label{p6}
 \hat{P}_g=g\,\left(\begin{array}{cc}0&\mathbf \Gamma\\ \Gamma &
\mathbf M\end{array}\right)-dg^{-1}\,(M,\mathbf A\g)\left(\begin{array}{cc}0&0\\ 0&
\mathbf \Gamma\end{array}\right)\,.
\eq
Integrals of motion $(\g,M)$ and $(\g,\g)$ are the Casimir functions of $\hat{P}_g$ and two remaining integrals of motion
$H$ and $M^2$ are in involution with respect to the corresponding Poisson brackets. It allows us to get two commuting Hamiltonian vector fields
\[
v_1=\hat{P}_g\, \mathrm dH\,,\qquad v_2=\hat{P}_g\, \mathrm dM^2\,,
\]
and to apply the Lie integrability theorem after changing the time
$\mathrm dt\to g^{-1}\mathrm d t$ as proposed by Chaplygin \cite{ch03}.

It is easy to represent initial vector field $X$ (\ref{m-eq}) in the same form
\bq\label{x-ch}
X={g}^{-1}\, \widetilde{ P}_g\, \mathrm dH
\eq
using an almost Poisson bivector
\[
\widetilde{ P}_g=g\,\left(
 \begin{array}{cccc}
 0& 0 & 0& \boldsymbol{A} \\
 0 & 0 &0 & \boldsymbol{B} \\
 0 & 0 & 0 & \boldsymbol{\Gamma} \\
 \boldsymbol{A} & \boldsymbol{B} & \boldsymbol{\Gamma} & \boldsymbol{M} \\
 \end{array}
 \right)-dg^{-1}\,(M,\mathbf A\g)\left(
 \begin{array}{cccc}
 0& 0 & 0&0 \\
 0 & 0 &0 &0 \\
 0 & 0 & 0 & 0 \\
0 &0 & 0&\boldsymbol{\Gamma} \\
 \end{array}
 \right)\,.
\]
 We have to underline that initial vector field field $X$ ceases to be  a conformally Hamiltonian vector field on $\mathcal M$, dim$\mathcal M=12$ as bivector $ \widetilde{P} _g $ does not satisfy to the Jacobi identity and, thus, the field $ \widetilde{P} _g \, \mathrm dH $ is non-Hamiltonian.

Below we prove that there are several reductions of this almost-Poisson bivector to  Poisson bivectors.
Let us describe one partial example.

 \begin{prop}
 Bivector
\bq\label{pg-12}
{P}_g=\widetilde{P}_g+\left(
 \begin{array}{cccc}
 0& 0 & 0& \boldsymbol{A}_g \\
 0 & 0 &0 & \boldsymbol{B}_g \\
 0 & 0 & 0 & 0\\
 -\boldsymbol{A}_g^\top & -\boldsymbol{B}_g^\top & 0& 0 \\
 \end{array}
 \right)\,,
\eq
where
\[
\left(\boldsymbol{A}_g\right)_{ij}=(\alpha\times \g)_i\,\sigma_j\,,\qquad \left(\boldsymbol{B}_g\right)_{ij}=(\beta\times \g)_i\,\sigma_j
\]
and
\bq\label{f-fun}
\sigma_1=\dfrac{\g_1}{\g_1^2+\g_2^2}\,(1+g-\g_3\sigma_3)\,,\quad
\sigma_2=\dfrac{\g_2}{\g_1}\,\sigma_1\,,\quad
\sigma_3=\dfrac{d\g_3(a_1\g_1^2+a_2\g_2^2)}{d(a_1\g_1^2+a_2\g_2^2)-\g_1^2-\g_2^2}\,.
\eq
is a Poisson bivector. The Casimir functions of this Poisson bivector $P_g$, rank$P_g$=6\,,
\[
{P}_g\mathrm d C_k=0\,,\qquad k=1,\dots,6\,,
\]
coincide with geometric integrals $C_1,\dots,C_6$ (\ref{c-int}).
\end{prop}
The proof is a straightforward verification.

Unit vectors $\alpha,\beta$ and $\g$ form an inertial orthonormal frame, so we can rewrite entries of $\boldsymbol{A}_g$ and $\boldsymbol{B}_g$ in the following way
\[
\left(\boldsymbol{A}_g\right)_{ij}=-\beta_i\,\sigma_j\,,\qquad \left(\boldsymbol{B}_g\right)_{ij}=\alpha_i\,\sigma_j\,.
\]
\begin{rem}
Bivector ${P}_g$ is a Turiel type deformation of the canonical bivector $P$ (\ref{can-poi} and, therefore, there is a family of  Poisson maps which reduce ${P}_g$ to canonical form \cite{ts12a,ts12b}. We present only one partial mapping $M\to L$, which reduces $P_g$ to $P$ (\ref{can-poi})
\ben
L_1&=&g^{-1}\Bigl(M_1+\frac{(g-\g_3 \sigma_3-1)(\g,M)}{\g_1^2+\g_2^2}\,\g_1\Bigr)\,,\qquad L_3=g^{-1}\Bigl(M_3+\sigma_3(\g,M)\Bigr)\,,\nn\\
L_2&=&g^{-1}\Bigl(M_2+\frac{(g-\g_3 \sigma_3-1)(\g,M)}{\g_1^2+\g_2^2}\,\g_2\Bigr)\,,
\en
 so that
\[
(\g,M)=(\g,L)\,.
\]
According to \cite{ts12a,ts12b} there are other Poisson maps expressed via elliptic integrals, which do not have  singularity at $\g_{1,2}=0$.
\end{rem}

Using the  Poisson bivector $P_g$ (\ref{pg-12}) we can easily prove that the nine functionally independent integrals of motion
\[C_1,\dots,C_6\,,\qquad f_1=H\,,\qquad f_2= M^2,\qquad f_3= (\g,M)
\]
are in  involution with respect to the corresponding Poisson bracket $\{.,.\}_g$. Spatial com\-po\-nents $J_k$ (\ref{j-int}) of momenta have the following brackets
\[
\{J_2,J_3\}_g=J_1\,,\qquad \{J_1,J_3\}_g=-J_2\,,
\]
whereas bracket $\{J_1,J_2\}_g$ is not an integral of motion in contrast with the holonomic case (\ref{j-br}).

Three independent integrals of motion $f_k$ in involution give rise to linear independent commuting vector fields
\bq\label{u-ch}
u_2=P_g\mathrm dH\,,\qquad u_3=P_g\mathrm d M^2\,,\qquad u_4=P_g\mathrm d(\g,M)\,,\qquad [u_i,u_j]=0\,.
\eq
The initial vector field $X$ (\ref{m-eq}) is a linear combination
\bq\label{decomp-Ch}
X={g}^{-1}\, ( u_2-s_1\,u_4)\,,
\eq
of these Hamiltonian vector fields, where
\[
s_1=(a_1\g_1M_1+a_2\g_2M_2)\left(\frac{1-da_3\g_3^2}{g(\g_1^2+\g_2^2)}-\frac{1}{d(a_1\g_1^2+a_2\g_2^2)-\g_1^2-\g_2^2}\right)
+\frac{(a_1\g_1^2+a_2\g_2^2)da_3\g_3M_3}{g(\g_1^2+\g_2^2)}\,.
\]
Similar decompositions (\ref{decomp-Ch}) of the original vector field on the Hamiltonian vector fields for other nonholonomic systems are discussed in \cite{bts12,ts13, ts13a}.

 In contrast with the holonomic case in $d\neq 0$ vector field $X$ has only one symmetry field
 \[u_4=P_g\mathrm d (\g,M)\,,\qquad [X,u_4]=0\,,\]
 because
 \[ [X,u_{2}]\neq 0\,,\qquad [X,u_{3}]\neq 0.\]
 In order to get the missing symmetry vector field we can apply the Chaplygin change of time
\[\mathrm dt\to g^{-1}\mathrm d t\]
and define three independent vector fields
\bq\label{alg-v}
v_1=gX=u_2-s_1\,u_4\,,\qquad v_2=u_3-s_2\,u_4\,,\qquad v_3=u_4\,.
\eq
If coefficient $s_2$ satisfies the relation
\bq\label{eq-s}
\mathcal L_{u_2} s_2=\mathcal L_{u_3}s_1\,,\qquad \mathcal L_{u_4} s_2=0\,,
\eq
these vector fields $v_1,v_2$ and $v_3$ (\ref{alg-v}) generate an Abelian algebra with respect to the standard commutator $[. ,. ]$ .

After integration of (\ref{eq-s}) one gets an explicit expression for this coefficient
\ben
s_2&=&2(\g_1M_1+\g_2M_2)\left(\frac{d(a_1\g_1^2+a_2\g_2^2)-1}{d(a_1\g_1^2+a_2\g_2^2)-\g_1^2-\g_2^2}
+\frac{1-da_3\g_3^2}{g(\g_1^2+\g_2^2)}\right)\label{s2}\\
\nn\\
&+&\frac{2d\g_1\g_2(a_1-a_2)(\g_2M_1-\g_1M_2)}{g(\g_1^2+\g_2^2)}+2d\g_3M_3\left(\frac{a_1\g_1^2+a_2\g_2^2}{d(a_1\g_1^2
+a_2\g_2^2)-\g_1^2-\g_2^2}+\frac{a_3}{g}\right).\nn
\en
Thus, we constructed a necessary set of commuting vector fields. It is easy to prove that
\[\mathcal L_{v_i} f_j = 0 \,,\qquad i,j=1,2,3\,,\]
and
\[\mathcal L_{v_1} J_{1,2}=\mathcal L_{v_2} J_{1,2}= 0\,.\]
Here $J_{1,2}$ the spatial components of momenta.

\begin{prop}
For the Chaplygin ball vector field $X$ (\ref{m-eq}) at 12-th dimensional phase space $\mathcal M$
we can adduce nine functionally independent integrals of motion
 \[C_1,\dots,C_6\,,\qquad f_1=H\,,\qquad f_2= M^2\,,\qquad f_3= (\g,M)\,,
\]
 and three linearly independent vector fields  commuting to each other's
 \[
v_1=gX=P_g\mathrm d f_1-s_1P_g\mathrm df_3\,,\qquad v_2=P_g\mathrm d f_2-s_2\,P_g\mathrm d f_3\,,\qquad v_3=P_g\mathrm d f_3\,,
\]
including original field after the time change.

Because the Lie derivatives along any of the fields
 $v_i $ from any of the integrals $ f_j $ are equal to zero, it is enough for $X$ to be  integrable by quadratures according to the Lie integrability theorem.
\end{prop}

It is a formal proof of integrability, the constructive procedure of integrability by quadratures has been proposed by Chaplygin \cite{ch03}, see also \cite{kil02} and references within.

\begin{rem}
By definition the  Hamiltonian vector fields $u_1,u_2$ and $u_3$ (\ref{u-ch}) are  Poisson vector fields for the Poisson bivector $P_g$
\[
\mathcal L_{u_i}P_g=0\,,\qquad i=1,2,3,
\]
 whereas their linear combinations are not
 \[
\mathcal L_{v_k}P_g\neq 0\,, \qquad k=1,2.
\]
Of course, since the vector fields $v_k$ pairwise commute, we have
\[\mathcal L_{v_k}(v_1\wedge v_2\wedge v_3)=0\,,\qquad k=1,2,3,\]
similar to the Hamiltonian mechanics \cite{van11}, so the  three equations
$\dot{x}=v_k$
can be simultaneously integrated by quadratures.
\end{rem}

\subsection{The Turiel type deformations of the canonical Poisson bivector}
Let
 $\theta,\psi,\phi$ be the Euler nutation, precession, and rotation angles associated with the fixed in  space ortonormal frame, respectively. So that the corresponding rotation  matrix is
$$
{\mathbf R}= \begin{pmatrix}
 \cos\phi\cos\psi - \cos\theta \sin\psi \sin\phi & \cos\phi\sin\psi + \cos\theta \cos\psi \sin\phi & \sin\phi \sin\theta \\
 -\sin\phi\cos\psi - \cos\theta \sin\psi \cos\phi & -\sin\phi\sin\psi + \cos\theta\cos\psi\cos\phi & \cos\phi \sin\theta \\
 \sin\theta \sin\psi & -\sin\theta \cos\psi & \cos\theta
\end{pmatrix}.
$$
Using the standard definition of  angular momenta via variables $p_\phi,p_\theta$ and $p_\psi$
\[\begin{array}{l}
M_1 =\left(\frac{\sin\phi}{\sin\theta}\,\bigl(\cos\theta\,p_\phi+p_\psi\bigr)-\cos\phi\,p_\theta\right)\,,\\
\\
M_2=\left(\frac{\cos\phi}{\sin\theta}\,\bigl(\cos\theta\,p_\phi+p_\psi\bigr)+\sin\phi\,p_\theta\right)\,, \qquad
M_3 = -p_\phi\,.
\end{array}
\]
we can rewrite the equations of motion (\ref{m-eq}) as equation on six dimensional phase space with coordinates ${x}=(\phi,\theta,\psi,p_\phi,p_\theta,p_\psi)$.

If we know solutions $\phi(t)$, $\theta(t)$, $p_\phi(t)$ and $p_\theta(t)$, we can obtain a
precession angle by quadrature
\begin{equation} \label{dot_psi} \dot \psi = \frac { \omega_1 \g_1 + \omega_2 \g_2}{g(\g_1^2 + \g_2^2)},
\end{equation}
see \cite{ch03}. As usual, momenta associated with the cyclic variable $\psi$ is a constant of motion
\[{p}_\psi=(\g,M)=const\]
that allows us to reduce the initial six dimensional phase space $\mathcal M$ to four dimensional phase $\hat{\mathcal M}$ space with coordinates $\hat{x}=(\phi,\theta,p_\phi,p_\theta)$.

This phase space $\hat{\mathcal M} $ is diffeomorphic to the cotangent bundle of the two-dimensional unit sphere $T^*S^2$. According to \cite{ts12a} there are  Turiel type deformations of the canonical Poisson bivector
\[
P=\left(
 \begin{array}{cc}
 0 & \mathrm I \\
 -\mathrm I & 0 \\
 \end{array}
 \right)
\]
depending on two arbitrary functions $g(\phi,\theta)$ and $h(\phi,\theta)$
 \bq\label{pg-sph}
 \hat{P}_g=\left(
 \begin{array}{cccc}
 0 & 0 & g(\phi,\theta) & 0 \\
 * & 0 & 0 &g(\phi,\theta) \\
 * & * & 0 & \partial_\theta g\,p_\phi-\partial_\phi g\,p_\theta- h(\phi,\theta)\\
 * & * &* & 0\\
 \end{array}
 \right)\,.
 \eq
 Here à $\partial _\phi g$ and $\partial_\theta g$ are derivatives of $g(\phi,\theta)$ by variables $\phi$ and $\theta$, respectively.

 In the particular case if $g(\phi,\theta)$ is given by (\ref{g-fun}) and
 \bq\label{w-ch}
 h(\phi,\theta)=w(\phi,\theta)p_\psi\,,\quad w(\phi,\theta)=g^{-1}\,d\sin\theta\,(a_1\sin^2\phi+a_2\cos^2\phi)\,,
 \eq
one gets bivector $\hat{P}_g$ (\ref{p6}), which was found by Borisov and Mamaev \cite{bm01}. Expression for $g(\phi,\theta)$ is fixed by invariant volume form $\Omega$ (\ref{volume-form}), whereas function $h(\phi,\theta)$ is uniquely defined by the involutivity condition
\[\{H,M^2\}_g=0\,.\]

Let us consider lifting  the Poisson bivector $\hat{P}_g$ with arbitrary $g(\phi,\theta)$ and $h(\phi,\theta)=w(\phi,\theta)p_\psi$ to the rank-six Poisson bivector on the initial phase space $\mathcal M$.
\begin{prop}
 Two arbitrary functions $g(\phi,\theta)$ and $w(\phi,\theta)$ define the Poisson bivector with linear in momenta entries
\bq\label{pg-sph2}
\tilde{P}_g=\left(
 \begin{array}{cccccc}
 0 & 0 & 0 & g(\phi,\theta) & 0 & 0 \\
 * & 0& 0 & 0 & g(\phi,\theta) & 0 \\
 * & * & 0 & u(\phi,\theta) & v(\phi,\theta) & 1 \\
 * & * & * & 0 & \partial_\theta gp_\phi-\partial_\phi gp_\theta- w(\phi,\theta)p_\psi & 0 \\
 * & * & * & * & 0 & 0 \\
 * & * & * & * & * & 0 \\
 \end{array}
 \right)\,,
\eq
 on the six dimensional phase space $\mathcal M$ with coordinates ${x}=(\phi,\theta,\psi,p_\phi,p_\theta,p_\psi)$,
if
\bq\label{uv-ch}
g\Bigl(\partial_\phi v(\phi,\theta)-\partial_\theta u(\phi,\theta)\Bigr)+u(\phi,\theta)\,\partial_\theta g-v(\phi,\theta)\,\partial_\phi g-w(\phi,\theta)=0\,.
\eq
Bivector $P_g$, rank $P_g$=6, defines a symplectic form and an invariant volume form with  density $g^{-1}(\phi,\theta)$ on the phase space $\mathcal M$, which becomes a symplectic manifold .
\end{prop}
The proof is straightforward.

\begin{rem}
Change of variables
\[p_\phi=g^{-1}(p_\phi-u(\phi,\theta)p_\psi)\,,\quad\mbox{and}\quad p_\theta=g^{-1}(p_\theta-v(\phi,\theta)p_\psi)\]
reduces bivector $P_g$ to the canonical bivector
\[
P=\left(
 \begin{array}{cc}
 0 & \mathrm I \\
 -\mathrm I & 0 \\
 \end{array}
 \right)\,.
\]
We can say that $P_g$ is a deformation of $P$ in the Poisson-Lichnerowicz cohomology \cite{ts11}.
\end{rem}

For the Chaplygin ball $g(\phi,\theta)$ and $w(\phi,\theta)$ are given by (\ref{g-fun}) and (\ref{w-ch}), respectively. In this case there is a "simple" partial solution of equation (\ref{uv-ch})
\[
 v(\phi,\theta)=0,\qquad
u(\phi,\theta)=gF(\phi)+\dfrac{d\cos\theta(a_1\sin^2\phi+a_2\cos^2\phi)}{1-da_1\sin^2\phi-da_2\cos^2\phi}\,,
\]
where $F(\phi)$ is an arbitrary function. At $F(\phi)=0$ this particular choice of functions $g$, $w$, $v$ and $u$
corresponds to the bivector $P_g$ (\ref{pg-12}).

Let us show another solution for the same $g(\phi,\theta)$ and $w(\phi,\theta)$
\ben
 v(\phi,\theta)&=&\frac{g}{\sin\theta\sqrt{1-da_1\sin^2\theta-da_3\cos^2\theta}}\Bigl(E_1(\cos\phi,z)+\Bigr.\nn\\
 &&\qquad\qquad\qquad+\left.
\frac{ \mathrm i(1-da_3\cos^2\theta)E_3(z\cos\phi,1,z^{-1})}{\sin\theta\sqrt{d(a_1-a_2)(1-da_1\sin^2\theta-da_3\cos^2\theta)\,}}\right)\,,\nn
\\
u(\phi,\theta)&=&0\,. \nn
\en
Here $E_1$ and $E_3$ are incomplete elliptic integrals of the first and  third kind, respectively, and
\[
z=\frac{\mathrm i\sin\theta\sqrt{d(a_1-a_2)\,}}{\sqrt{1-da_1\sin^2\theta-da_3\cos^2\theta}}\,,\qquad \mathrm i=\sqrt{-1}\,.
\]
Of course, there are a lot of other solutions of the equation (\ref{uv-ch}) for the given $g(\phi,\theta)$ and $w(\phi,\theta)$.

\begin{rem}
We can not put $s_1=0$ in the decomposition (\ref{decomp-Ch}) using some special choice of the functions $u(\phi,\theta)$ and $v(\phi,\theta)$.
\end{rem}

\subsection{Rank-two Poisson bivectors associated with the Chaplygin ball}
The Lie symmetry analysis provides an algorithm to determine a  set of infinitesimal symmetries associated with the given equations of motion. Finding the Jacobi last multiplier and integrals of motion from the Lie symmetries is often a cumbersome procedure. On the other hand it is very easy to get rank-two Poisson structures directly from symmetries \cite{mr}.

Namely, for the Chaplygin ball we have two families of pairwise commuting vector fields
\[
u_1=P_g\mathrm d f_1\,,\qquad u_2=P_g\mathrm d f_2\,,\qquad u_3=P_g\mathrm d f_3\,,\qquad [u_i,u_j]=0
\]
and
\[
v_1=u_1-s_1u_3\,,\qquad v_2=u_2-s_2u_3\,,\qquad v_3=u_3\,,\qquad [v_i,v_j]=0 .
\]
Using these vector fields we can construct two families of  bivectors
\[
P_u^{(i.j)}=u_i\wedge u_j\,,\qquad\mbox{and}\qquad P_v^{(i.j)}=v_i\wedge v_j\,,\qquad i,j=1,2,3.
\]
It is easy to prove that the Schouten bracket for these bivectors are equal to zero, for instance
\[
[P_v^{(i,j)},P_v^{(i,j)}]=2v_i\wedge v_j\wedge [v_i,v_j]=0\,.
\]
It means that we have two families of the Poisson bivectors having a constant rank two,  see   \cite{mr}. Entries of these bivectors are the fourth order polynomials
in momenta whereas entries of initial Poisson bivector $P_g$ are linear polynomials.

Associated with the Poisson bivector $P_v^{(i,j)}$ brackets  is equal to
\[
\{F,G\}_v^{(i,j)}=(\mathrm d F\wedge \mathrm d G, P_v^{(i,j)})=\mathcal L_{v_i} F\cdot \mathcal L_{v_j}G-
\mathcal L_{v_i} G\cdot \mathcal L_{v_j}F\,.
\]
Sequently,  integrals of motion $f_1,f_2$ and $f_3$ are  the Casimir functions for these Poisson brackets because
\[
\mathcal L_{v_k} f_m=0\,,\qquad \forall k,m=1,2,3.
\]
 In fact  we have two families of compatible Poisson bivectors, i.e. linear combinations
\[
P_u=\lambda\,P_u^{(1,2)}+\mu\,P_u^{(1,3)}+\nu\,P_u^{(2,3)}\,,\qquad\mbox{and}\qquad
P_v=\lambda\,P_v^{(1,2)}+\mu\,P_v^{(1,3)}+\nu\,P_v^{(2,3)}
\]
are also the Poisson bivectors for any $\lambda,\mu$ and $\nu$.

In \cite{bts12} we used linear combinations of the rank-two Poisson bivectors in order to get rank-four Poisson bivectors for the nonholonomic  Routh sphere. It will be interesting to study similar linear combinations in the Chaplygin case too.

\section{Possible generalisations}
 \setcounter{equation}{0}
 Using functionally independent integrals of motion in involution we construct a set of commuting Hamiltonian vector fields \[u_k=P\mathrm d f_k\,,\qquad [u_i,u_j]=0\,,\]
 which form a basis with a necessary number of generators for applicability of the Lie integrability theorem. Other combinations could be taken
 \bq\label{v-gen}
 v_i=\sum s_{ik}u_k\,,
 \eq
 in order to include the original vector field $v_1=gX$ and the necessary number of symmetries into the abelian (solvable) algebra. Coefficient $s_{ik}$ may be obtained by  direct integration of a system of equations similar to (\ref{eq-s}).

\begin{rem}
Recall that in the toric integrability theory we also consider special linear combinations of the initial commuting Hamiltonian vector fields in order to get the so-called fundamental vector fields of the torus action \cite{neh05,van11}.
\end{rem}

It is easy to see that in this direct method we do not use the known volume form $\Omega$ (\ref{volume-form}), which is a key element of the Euler-Jacobi integrability theorem. So, we hope that there is another constructive method of  finding  vector fields $v_k$ (\ref{v-gen}) using the volume form, its relation with symmetries (\ref{mu-det}) , torus action theory, modular vector fields or something else.

\subsection{Motion in the Brun potential field}
 Let us consider the motion of the Chaplygin ball in the so-called Brun potential field \cite{koz85}. In order to get the corresponding vector field $X_b$ we have to add a momenta of external force to the equation for the angular momenta in (\ref{m-eq})
\bq\label{eq-brun}
M=M\times \omega+\frac{\partial U}{\partial \g}\times \g\,,\qquad U=c(a_2,a_3\g_1^2+a_1a_3\g_2^2+a_1a_2\g_3^2)\,,
\eq
 In this case we have nine integrals of motion $C_1,\ldots,C_6$ and
\bq\label{int-brun}
 f_1=\dfrac12\,(M,\omega)+U,\quad f_2=M^2-2c(a_1\g_1^2+a_2\g_2^2+a_3\g_3^2),\quad f_3=(\g,M)
\eq
and the same invariant volume form (\ref{volume-form}). However, functions $J_1=(\alpha,M)$ and $J_2=(\beta,M)$ are no longer constants of motion and we cannot apply the Euler-Jacobi integrability theorem in this case.

Of course, equations for $\g$ and $M$ on the six dimensional phase space $\hat{\mathcal M}$ are integrable by quadratures according to the Euler-Jacobi theorem, and, after integration of these equations of motion for $\g$ and $M$, we can integrate equations for the remaining variables $\alpha$ and $\beta$ by quadratures as well. But we do not have a formal proof of integrability on 12-th dimensional phase space $\mathcal M$.

According to \cite{ts11}, vector filed $\hat{X}_b$ on the six-dimensional phase space $\hat{\mathcal M}$ is a conformally Hamiltonian vector field
\[
\hat{X}_b={g}^{-1}\, \hat{P}_g\, \mathrm d(H+U)
\]
where $\hat{P}_g$ is the same Poisson bivector (\ref{p6}) as for the free Chapligin ball.

The lifting of the Poison bivector $\hat{P}_g$ on the 12-dimensional phase space is given by (\ref{pg-12}). Using this Poisson bivector and integrals of motion we can construct three independent commuting vector fields
\[
u_2=P_g\mathrm df_1\,,\qquad u_3=P_g\mathrm d f_2\,,\qquad u_4=P_g\mathrm df_3\,,\qquad [u_i,u_j]=0\,.
\]
The original vector field $X_b$ (\ref{m-eq},\ref{eq-brun}) is a linear combination
\bq\label{decomp-brun}
X={g}^{-1}\, ( u_2-s_1\,u_4)\,,
\eq
of these Hamiltonian vector fields, where $s_1$ is the same coefficients as in (\ref{decomp-Ch}).

 As in the previous Section we can apply the Chaplygin change of time
and define the three independent commuting vector fields
\[
v_1=gX_b=u_2-s_1\,u_4\,,\qquad v_2=u_3-(s_2+ \delta s)\,u_4\,,\qquad v_3=u_4\,.
\]
Here $s_2$ is given by (\ref{s2}) and $\delta s$ is the solution of the following differential equation
\bq\label{seq-brun}
\{f_1,\delta s\}=4cd(a_1-a_2)(a_2-a_3)(a_3-a_1)\,\g_1\g_2\g_3\,,
\eq
where $f_1$ is  mechanical energy. We do not have  an explicit observable solution of this equation.

\subsection{Motion of the contact point}
The trajectory of the contact point on the plane is also important for understanding the
ball's motion in  absolute space.  The equations of motion of the contact point can be obtained using the condition
that its velocity is equal to zero in the body frame
\[
v=r\times\omega=0\,.
\]
Here $v$ is the velocity of the center of mass, $r = b\gamma$
 is a vector, connecting the center of mass with the contact point. In terms of projections on the axes of coordinates
 this condition looks like
 \bq\label{eq-xy}
\dot{x}=(v,\alpha)=b (\omega,\beta)\,,\qquad \dot{y}=(v,\beta)=-b(\omega,\alpha)\,.
\eq
 Here $x$ and $y$ are  coordinates of the contact point on the plane and $b$ is the ball's radius \cite{ch03,kil02}.

Thus, on the phase space $\mathcal M_{c}$, dim$\mathcal M_{c}=14$ we have a vector field $X_{c}$
defined by equations of motion (\ref{m-eq}) and (\ref{eq-xy}). As above, we can try to prove the integrability of this vector field using the Poisson geometry methods.

\begin{prop}
By adding Poisson brackets between new variables
\[
\{x,y\}_g=1\,,
\]
to the Poisson brackets $\{.,.\}_g$ (\ref{pg-12}) one gets a rank eight Poisson bivector $P_{c}$ on the $14$-th dimensional phase space with the Casimir functions $C_1,\ldots,C_6$.
\end{prop}
In this case the 14-th dimensional vector field $X_{c}$ is a linear combination of the Hamiltonian vector fields
\bq\label{x-cont}
{X}_{c}=g^{-1}\Bigl(P_{c}\mathrm d f_1- s_1P_{c}\mathrm d f_3-s_2P_{c}\mathrm d x-s_3P_{c}\mathrm dy)\Bigr)\,,
\eq
 which has two symmetries associated with the cyclic variables $x$ and $y$. Here
 \[
 s_2=bg(\omega,\beta)\quad\mbox{and}\quad s_3= bg(\omega,\alpha)\,.\]
So, we have nine integrals of motion in involution $C_1,\ldots,C_6,f_1\,\ldots,f_3$
and five commuting vector fields
\[
u_1=P_{c}\mathrm d f_1\,,\quad u_2=P_{c}\mathrm d f_2\,,\quad u_3=P_{c}\mathrm d f_3\,,\quad u_4=P_{c}\mathrm d x\,,\quad u_5=P_{c}\mathrm d y\,.
\]
After the change of time the initial vector field $v_1=gX_c$ commutes with only two symmetry fields
\[
[v_1,u_4]=[v_1,u_5]=0\quad\mbox{and}\quad[v_1,u_k]\neq 0\,,\quad k=1,2,3\,.
\]
Therefore, we have to add two linear combinations of $u_k$
\bq\label{s-cont}
v_k=\sum_{j=1}^5 s_{kj}u_j\,,\qquad k=2,3.
\eq
to the  field $v_1$ and its symmetries $v_4=u_4,\,v_5=u_5$ associated with cyclic variables $x,y$.
Here $s_{kj}$ are some locally defined functions on phase space.

Supposing that $v_1,\ldots,v_5$ generate a solvable (nilpotent or Abelian) algebra one gets a system of differential equations on $s_ {kj} $ (\ref{s-cont}) similar to (\ref{eq-s}) and (\ref{seq-brun}), which we have to integrate.

In some sense we are looking for an inversion of the Routh procedure, which in its original form \cite{rou55}, was concerned with eliminating  the generalized velocities corresponding to ignorable
or cyclic coordinates from a Lagrangian problem. The modern discussion of the Routh procedure and its inversion based on the notion of  anholonomic frames may be found in \cite{cr08}.

We would like to thank Alexey Borisov and Ivan Mamaev for  genuine interest and helpful discussions.
This work was partially supported by RFBR grant 13-01-00061.

\end{document}